# Spatially resolved, energy-filtered imaging of core level and valence band photoemission of highly p and n doped silicon patterns

**N Barrett**[1], **L F Zagonel**[1], **O Renault**[2] **and A Bailly**[2]

[1] CEA DSM/IRAMIS/SPCSI, CEA Saclay, 91191 Gif sur Yvette, France
[2] CEA-LETI, MINATEC, 17 rue des Martyrs, 38054 Grenoble Cedex 9, France   E-mail: nick.barrett@cea.fr

**Abstract**
An accurate description of spatial variations in the energy levels of patterned semiconductor substrates on the micron and sub-micron scale as a function of local doping is an important technological challenge for the microelectronics industry. Spatially resolved surface analysis by photoelectron spectromicroscopy can provide an invaluable contribution thanks to the relatively non-destructive, quantitative analysis. We present results on highly doped n and p type patterns on, respectively, p and n type silicon substrates. Using synchrotron radiation and spherical aberration-corrected energy filtering, we have obtained a spectroscopic image series at the Si 2p core level and across the valence band. Local band alignments are extracted, accounting for doping, band bending and surface photovoltage.

## 1. Introduction

The increasing integration of semiconductor devices makes the availability of reliable, quantitative, non-destructive analysis tools more and more important. Electron microscopy is well suited to the task in terms of spatial resolution. However, in order to obtain quantitative information on the chemistry and the electronic structure, the electron energies used must be reduced to a few hundred eV in order to privilege in-plane surface and near surface contrast with respect to bulk heterogeneity. X-ray photoelectron emission microscopy (XPEEM) offers precisely the possibility to combine the required spatial and chemical state resolution. In particular, by using soft x-ray excitation combined with a full energy-filtered analysis, one may carry out XPEEM spectromicroscopy, combining the elemental, chemical and electronic structure sensitivities of classical photoelectron spectroscopy (PES) with spatial resolutions on the 100 nm scale, now easily accessible with current PEEM instruments. Image series as a function of photoelectron kinetic energy $E_k$ are acquired step by step over the energy window of interest. Each image gives the intensity as a function of position within the microscope field of view (FoV). A full image series is a three-dimensional (3D) dataset $I(x, y, E_k)$ allowing extraction of photoemission spectra from any area of interest (AOI) within the field of view.

One particularly important application is the study of the electronic structure as a function of spatially dependent doping levels of silicon, whether in the form of doped patterns or gradients. Thus, whereas gate thicknesses are of the order of 1 nm, lateral device dimensions are closer to 100 nm. Although the dopant concentration is usually well below the PES detection levels, the effects on the band alignments should be readily measured. In particular, shifts in the Fermi level position as a function of doping, and band bending at the surface or at an interface, should be measurable.

As Phaneuf et al [1] observed, this is a four-dimensional problem. Laterally, the electronic energy levels $E$ are imaged as a function of position $(x, y)$, including band bending across pn junctions, whereas perpendicular to the surface band bending has to be included as a function of $z$ (whether this be due to the surface termination or to the presence of an oxide layer over the silicon substrate). Finally, there may also be unwanted additional effects such as surface contamination, either intrinsic to the wafers or as a result of x-ray irradiation during analysis, particularly when using undulator radiation at a synchrotron source.

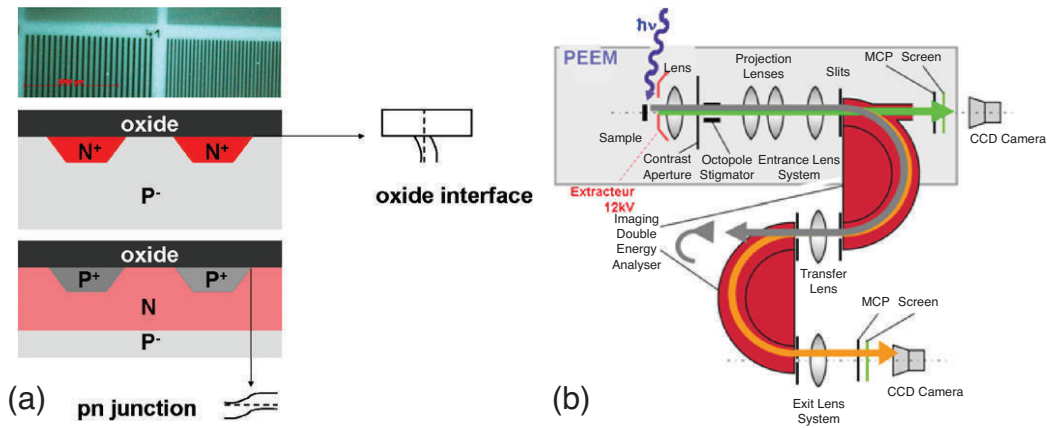

**Figure 1.** (a) Top: optical micrograph of doped patterns. Middle: $N^+/P^-$ sample. Bottom: $P^+/N$ sample. Also shown: expected band line-ups in $x$–$y$ (pn junction) and $z$ (oxide/substrate interface). (b) Schema of the XPEEM instrument used showing the direct PEEM, selected-area XPS and energy-filtered PEEM modes. The double hemispherical analyser suppresses the $\alpha^2$ aberrations inherent in a single hemispherical analyser.

The first observation of contrast between p and n doped regions using electron microscopy was done by Chang and Nixon [2] using secondary electrons in a scanning electron microscope. In further studies by other groups, the basic rule was that p type areas were brighter than n type areas. This was attributed to the difference in band bending at the surface as a function of dopant type and concentration, and some sophisticated models have been developed [3]. However, the use of secondary electrons for imaging, or the use of UV excitation sources in PEEM experiments can make the extracted spectra extremely complex to disentangle. Frank *et al* [4] have recently reviewed the different contrast mechanisms which may play a role, including work function, band bending, electron–hole pair creation by hot electron absorption, direct transitions, surface contamination and spectral width of the photon source. Thus, the qualitative image contrast may be understood and related to the pn structures, but more quantitative analysis is extremely difficult. Furthermore, they did not benefit from full energy filtering. The image intensity thus represents integration over the band pass of the microscope, convoluted with the initial photon energy width. This has been partly overcome in PEEM experiments using a retarding energy filter [5]. Energy-filtered PEEM combined with a tuneable photon source such as synchrotron radiation can provide further information. It has already been used to study semiconductor surfaces, see for example [6, 7]. However, the most thorough study yet published [1] used a scanning photoelectron microscope to study variations in the electronic levels on a doped, patterned silicon wafer, thus combining lateral and energy resolution.

Here we present the first thorough fully energy-filtered XPEEM study of a similar system, which benefits both from the insights obtained by Phaneuf *et al* and the instrumental improvements by using a PEEM instrument optimized for spectromicroscopy. First the sample preparation and the XPEEM instrument used are described. Then we present the core level and valence band spectra extracted from regions of interest corresponding to different n and p type doping levels on the silicon wafers. The interpretation of the experimental results requires taking into account all of the contributions to the binding energies as measured by photoemission.

## 2. Experimental details

The samples were made in the Laboratoire d'Electronique et de Technologie de l'Information (LETI) at the CEA-Grenoble. The samples consisted of two-dimensional doped patterns implanted into silicon in the form of lines with variable spacing and width from 100 to 0.1 $\mu$m, see figure 1(a). Each set of lines was identified by figures with the same doping level, giving suitable shapes for optimization of the electron optics of the PEEM instrument. Two samples were studied, figure 1(a). The first sample, designated $N^+/P^-$, had $N^+$ doped zones ($10^{20}$ cm$^{-3}$) implanted into a $P^-$ substrate ($10^{16}$ cm$^{-3}$). The second, designated $P^+/N$, had $P^+$ patterns ($10^{20}$ cm$^{-3}$) implanted into an N type ($10^{17}$ cm$^{-3}$) substrate; the latter was itself deposited on a Si $P^-$ wafer ($10^{16}$ cm$^{-3}$). A native oxide layer was present on both samples. Figure 1(a) also illustrates the four-dimensional problem of band line-ups in such samples.

The spectromicroscopy experiments were done using a NanoESCA (Omicron Nanotechnology) [8] XPEEM instrument temporarily installed on the CIPO beamline of the ELETRRA synchrotron radiation source. A schema of the instrument is shown in figure 1(b). The NanoESCA has been specially designed to optimize spectral extraction in photoemission mode from small areas of interest. One major obstacle for most PEEM instruments is the transmission function. The transmission depends on the phase space acceptance of the instrument, which is in turn inversely proportional to the square of the pass energy when using a hemispherical-based energy analyser. Thus away from the photoemission threshold, the PEEM transmission sharply decreases, making imaging in photoemission mode difficult. Indeed, most spectroscopic PEEM work is done very efficiently by using the absorption mode at threshold and measuring the intensities as one sweeps through an absorption edge. The most systematic use of PEEM with full energy filtering is the work of Heun, Locatelli *et al* using a very bright photon source at the ELETTRA synchrotron, see for example [9, 10]. However, transmission remains a problem, and the energy resolution must often be relaxed to obtain good counting statistics [6]. The NanoESCA goes some way to overcoming this problem by an all electrostatic lens system and a low PEEM column voltage, close to the pass energy used in the energy analyser. Furthermore, the unique use of a double hemispherical analyser corrects the spherical $\alpha^2$ aberrations inherent in a single hemisphere, where $\alpha$ is the entrance angle of the electrons in the analyser [8]. Thus one may use pass energies and energy analyser slits comparable to those used in standard XPS without overly sacrificing phase space and thus transmission. This can be at the expense of the ultimate lateral resolution, but the system does permit simultaneous XPEEM imaging of core and valence levels in reasonable acquisition times.

The incident photon energy was 127 eV, representing the best compromise between the beamline undulator response and surface sensitivity. The PEEM contrast aperture diameter was 150 $\mu$m; a trade-off between lateral resolution and transmission. The double hemispherical analyser pass energy of the NanoESCA was 50 eV and 0.5 mm entrance slits were used. Image series through the Si 2p core level were acquired in 0.1 eV steps.

The overall estimated energy resolution was 0.15 eV, with a lateral resolution of 120 nm. The incident photon flux on the sample was rather low for synchrotron-based PEEM experiments ($\sim 10^{15}$ ph s$^{-1}$ cm$^{-2}$). The sample was close to ground potential, and an extraction voltage of 12 kV was used [11]. All results were taken using a 25 $\mu$m field of view. Image series as a function of the photoelectron kinetic energy were acquired for the Si 2p core level and for the valence band. The acquisition time per image for the former was 3 min, for the latter, 5 min. The photoelectron energy $E$ is measured with respect to the sample Fermi level $E_F$, thus $E - E_F = E_k + \Phi_{WF}$, where $E_k$ is the photoelectron kinetic energy and $\Phi_{WF}$ the sample work function. The variation of the work function under undulator radiation was also measured, and will be detailed in a forthcoming paper. The maximum intrinsic energy dispersion in the vertical direction of the FoV is less than 35 meV and the binding energy scale was calibrated using the Fermi level and the 3d emission of a flat polycrystalline sputtered Ag sample with exactly the same energy filter settings as for the doped silicon.

## 3. Results and discussion

### 3.1. Si 2p core level imaging

Figure 2 shows energy-filtered threshold images from the two samples. One can clearly distinguish the heavily doped

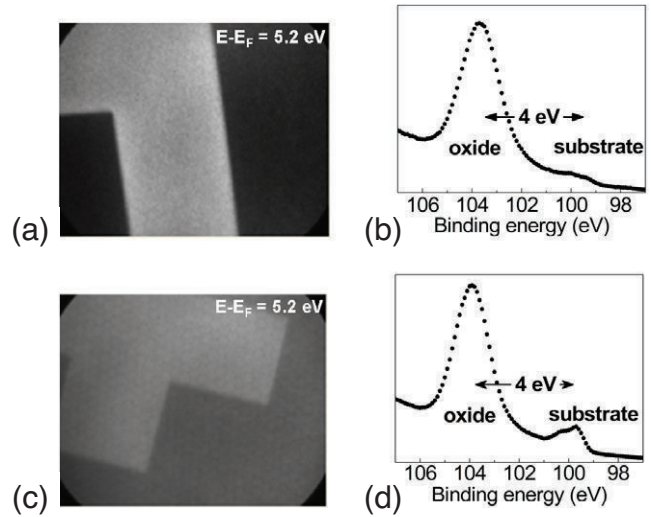

**Figure 2.** (a) and (c) Threshold images of N$^+$/P$^-$ and P$^+$/N, respectively, using a 25 $\mu$m field of view; (b) and (d) Si 2p spectra averaged over the field of view of (a) and (c), showing broad peaks given the energy resolution. The threshold images have been obtained using high resolution $1 \times 1$ binning.

N$^+$(P$^+$) regions on the P$^-$(N) substrate for N$^+$/P$^-$ (P$^+$/N). The image series show clear contrast inversion for both samples, and will be analysed in a forthcoming paper. On the right of figure 2 we show the area-averaged Si 2p spectra acquired on N$^+$/P$^-$ and P$^+$/N using the NanoESCA in the XPS mode. In this mode the instrument functions as a spectrometer using only the first hemispherical analyser which projects the photoelectrons onto a channeltron detection system. The spectra have two main peaks, one due to the native oxide at a binding energy of 104 eV and another, strongly attenuated, from the underlying bulk Si substrate at 100 eV. The separation of the two peaks, about 4 eV, corresponds to what one would expect for oxidized silicon [12]. However, given the energy resolution ($\sim 0.15$ eV) the spectral features are too large to be attributed to single peaks. Thus it may well be that the averaging of the photoemission signal from doped and undoped regions gives a broad signal, hiding the spatially dependent band alignments. This is a first indication that there is mixing of the electronic levels in the same field of view.

Figure 3 shows five images from the series of 91 images acquired through the Si 2p core level for the N$^+$/P$^-$ sample, with the corresponding binding energies (BE). This results in an image stack of the photoelectron intensity $I(x, y, E_k)$. At high binding energy the contrast inversion over the Si$^{4+}$ peak from the N$^+$ and P$^-$ doped regions is clearly seen, indicating visually that the Si$^{4+}$ peaks are shifted as a function of the doping type and concentration. There is a similar but much weaker inversion for the substrate peak. The small squares represent the areas of interest (AOIs) defined for spectral extraction ($400 \times 400$ nm$^2$). The resulting core level spectra are displayed in the bottom of figure 3 together with the best least square fits to the data. The colour code identifies the AOIs from which the spectra were extracted.

Standard core level fitting criteria were applied to the extracted spectra. After subtraction of a Shirley background function, a standard, five component deconvolution was applied, simulating the Si$^0$, Si sub-oxides and Si$^{4+}$ components [13]. Gaussian line shapes with a 2p spin–orbit splitting of 0.61 eV and a 2:1 branching ratio were used. The full width half maxima used for the Si$^0$, Si$^+$, Si$^{2+}$, Si$^{3+}$ and Si$^{4+}$ were 0.6, 0.9, 1.0, 1.1 and 1.4 eV, respectively. The experimental intensities extracted from AOIs distributed horizontally across the FoV are remarkably constant; the root mean square deviation being less than 5%. The fits to the experimental spectra are very good. The $\sim 2 \times 10^2$ counts per binned pixel give a count rate from each AOI of $\sim 2 \times 10^3$. The point $\chi^2$ residual between the experimental data and the fit is 2.2 ($\sim 0.1$%) for spectra from the $N^+$ regions and 3.0 for spectra from the $P^-$ regions. The standard deviation in the binding energy values obtained from spectra from different AOIs was 30 and 50 meV for the $N^+$ and $P^-$ regions, respectively.

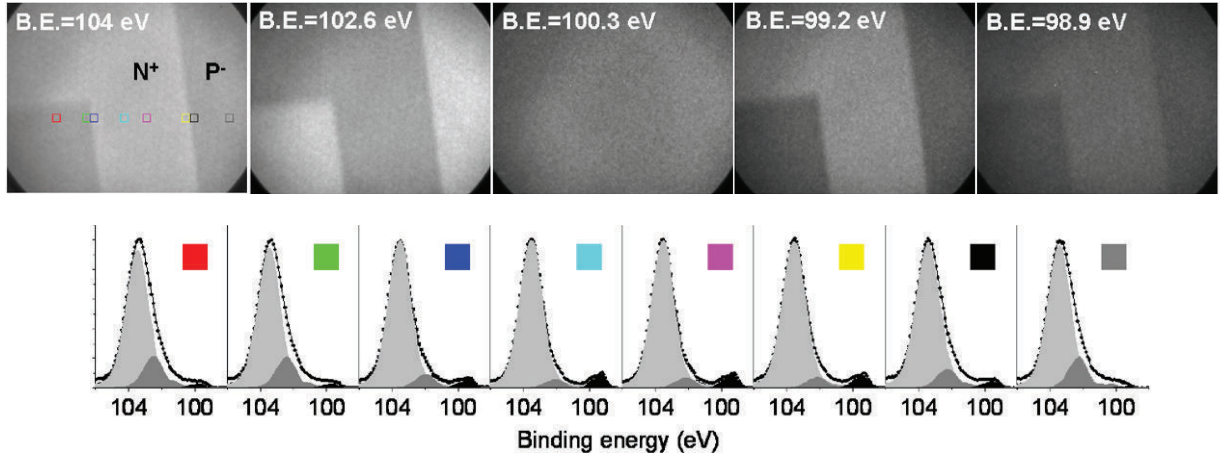

**Figure 3.** Top: selected images from the energy-filtered image series through the Si 2p core level on sample $N^+/P^-$ with a field of view of 25 $\mu$m. The first image shows the $400 \times 400$ nm$^2$ AOIs selected for extracting the local core level spectra. The contrast levels have been adjusted for better visibility. Bottom: core level spectra and best least square fits using standard literature parameters for Si 2p. The Si$^{4+}$ component due to the native SiO$_2$ oxide is in light grey, the sub-oxide components in grey and the Si$^0$ substrate component in black. The modulation of the oxide thickness as a function of doping is clearly visible. The variation in the absolute intensity of the spectra is less than 5% across the different AOIs.

Subtle changes occur in both the positions and the internal weightings of the Si 2p components. In particular, it is immediately obvious that the substrate signal is more strongly attenuated over the $P^-$ substrate than over the $N^+$ pattern, indicating that the native oxide thickness is dependent on the underlying doping level and type. Using the core level intensities and applying a simple model of exponential attenuation of the substrate signal due to the finite electron inelastic mean free path $\lambda$ [13] we can estimate the thickness of the oxide layer. Using the values of Himpsel et al [13] for the photoionization cross sections of Si$^0$ and Si$^{4+}$ at $h\nu = 130$ eV and the mean free path of the Si 2p electrons (0.75 nm) the $N^+(P^-)$ oxide thickness is 1.81 (2.60) nm. A smaller mean free path of 0.5 nm gives oxide thicknesses of 1.36 (1.90) nm respectively, and setting $\lambda_{Si} = \lambda_{SiO_2} = 0.35$ nm gives 1.07 (1.45) nm. It has been shown that for low $E_k$, the elastic component of the mean free path becomes significant due to scattering from final state band gaps [14]. Thus, the absolute value of the oxide overlayer thickness is uncertain. However, the oxide thickness difference $\Delta z$ over $N^+$ and $P^-$ regions, based on Himpsel et al's figures, can be calculated to be 0.79 nm. Mean free path values of 0.5 and 0.35 nm give $\Delta z$ equal to 0.54 and 0.38 nm, respectively.

We have also performed spectroscopic imaging of $N^+/P^-$ using an optimized, monochromatic laboratory Al K$\alpha$ source ($h\nu = 1486.7$ eV) [11]. At this energy $\lambda = 2.96$ nm, giving an average oxide thickness of 0.95 nm.

Figure 4 shows selected images from the Si 2p series of the $P^+/N$ sample, the definition of the AOIs and the extracted Si 2p spectra. In contrast to the $N^+/P^-$ spectra of figure 3, very little variation in the substrate Si$^0$ signal is observed as a function of doping type.

The two samples are not symmetric in their preparation. The $N^+$ patterns of $N^+/P^-$ are obtained by implantation of phosphor into a $P^-$ substrate, whereas the $P^+$ patterns of $P^+/N$ are obtained by boron implantation into a phosphor implanted layer (N) on a boron ($P^-$) implanted substrate. The strain fields are thus different in the two cases and $P^+/N$ has undergone a supplementary ion implantation to obtain the required doped patterns. Native oxide growth depends on the sample preparation history, it therefore seems reasonable that the oxide on $P^+/N$, having undergone successive ion implantations is more homogeneous than in $N^+/P^-$.

*3.2. Valence band imaging*

Image series were also acquired near the top of the valence band up to the Fermi level. The extracted spectra for $N^+/P^-$ and $P^+/N$ are shown in figure 5 with, for convenience, a reminder of the AOIs used. The Fermi level is marked by a dotted line. Two main features are visible in all spectra. Firstly, the O 2p orbitals which as expected peaked at 6–7 eV below the Fermi level, secondly, a small but significant contribution from the Si 3p orbitals, which comes from the underlying substrate. The Si 3p valence band contribution is attenuated the band bending between the substrate and the native oxide will extend over several nm, approximately $8L_D$, even for the most heavily doped patterns. Given the small photoelectron inelastic mean free path (∼0.35–0.75 nm, see above), this means that the valence band maximum (VBM) measured from our extracted spectra will indeed measure the onsets for the fully bent bands, and that we are far from the flat band regime. In fact, thanks to the use of tuneable synchrotron radiation, we are able to probe the spatial distribution of band bending at the buried substrate/oxide interface. The core level and valence band maximum results for the two samples are summarized in table 1.

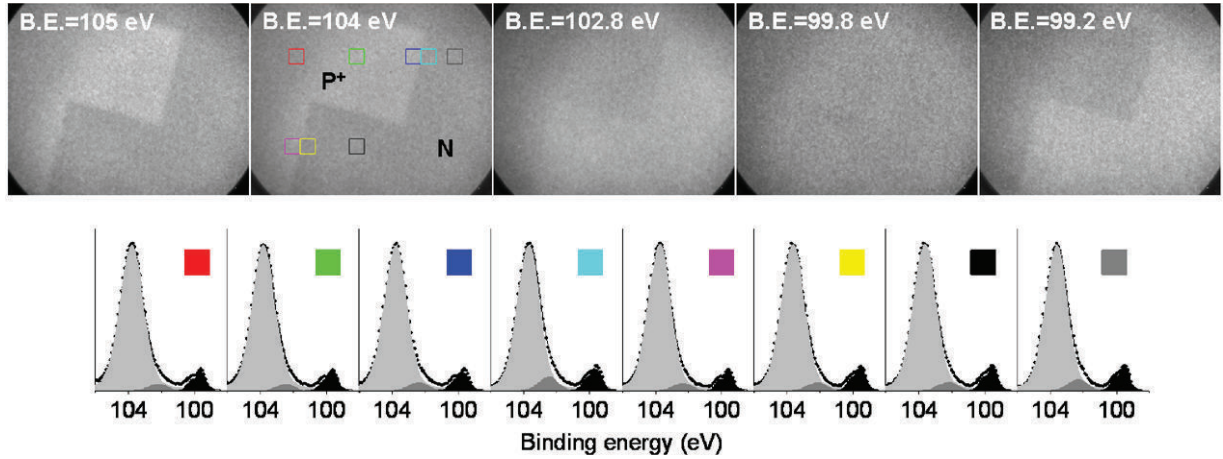

**Figure 4.** Top: selected binding energy images from the energy-filtered image series through the Si 2p core level on sample $P^+/N$ with a field of view of 25 $\mu$m. Contrast inversion through the $Si^0$ substrate peak is clearly visible. The second image shows the $400 \times 400$ nm$^2$ AOIs selected for extracting the local core level spectra. Bottom: core level spectra and best least square fits using standard literature parameters for Si 2p. The colour code corresponds to the AOIs defined on the second image. No modulation of the oxide thickness as a function of doping is visible.

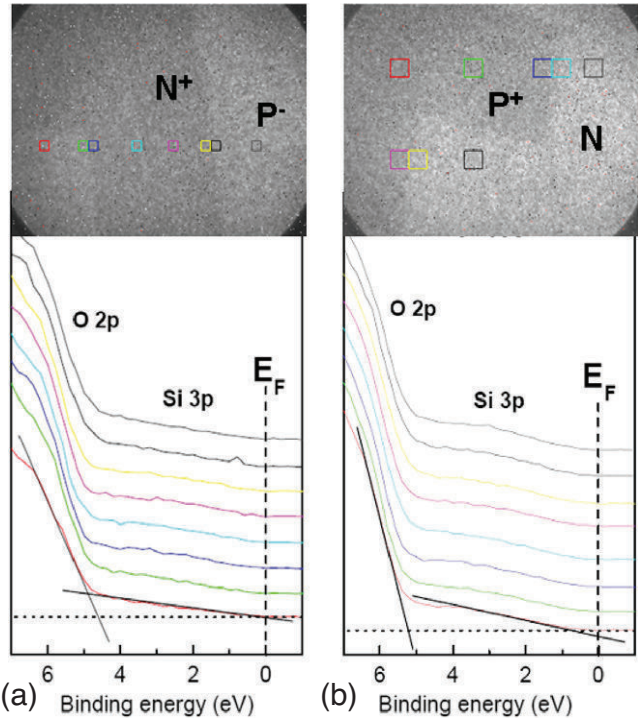

**Figure 5.** Top of the valence band spectra extracted from the same areas of interest as for the Si 2p core level spectra of figures 3 and 4. (a) $N^+/P^-$. (b) $P^+/N$. The densities of states due to the O 2p and Si 3p orbitals are marked, together with examples of linear extrapolation in order to determine the valence band onsets. The threshold images with the relevant AOIs are reproduced for convenience.

**Table 1.** Si 2p substrate binding energy, $Si^0/Si^{4+}$ core level shift, O 2p and Si 3p valence band onsets measured from the doped patterns and substrates of samples $N^+/P^-$ and $P^+/N$. All energies are in eV.

|  | $N^+/P^-$ | | $P^+/N$ | |
| --- | --- | --- | --- | --- |
|  | $N^+$ | $P^-$ | $P^+$ | N |
| $Si^0$ | $99.34 \pm 0.03$ | $99.60 \pm 0.05$ | $99.52 \pm 0.05$ | $99.59 \pm 0.03$ |
| $\Delta Si^{4+}$ | $4.02 \pm 0.03$ | $3.68 \pm 0.05$ | $4.06 \pm 0.05$ | $3.95 \pm 0.03$ |
| VBM O $2_p$ | 4.98 | 4.79 | 5.12 | 5.00 |
| VBM Si $3_p$ | — | — | 0.54 | 0.44 |

### 3.3. Band bending

The binding energy of a core electron in a doped semiconductor as measured by photoemission using an intense x-ray source can be written as:

$$BE = BE_{fb}^i + \Delta E_F^{doping} + \Delta E_{BB} + \Delta E_{SPV}, \quad (1)$$

where $BE_{fb}^i$ the binding energy is measured in the intrinsic case and in flat band conditions; $\Delta E_F^{doping}$ is the shift in the position of the Fermi level within the gap as a function of doping with respect to intrinsic silicon; $\Delta E_{BB}$ is the band bending at the by the native oxide, but is still visible. As for the Si 2p spectra, the attenuation is stronger in the spectra extracted from the p doped region than in those extracted from the n type pattern.

Extrapolating the upper edges of the O 2p and the Si 3p to zero allows a first determination of the valence band onsets. The Debye length is given by $L_D = \sqrt{\varepsilon_s kT/q^2 N}$ [16], thus substrate/oxide interface; and $\Delta E_{SPV}$ is the binding energy shift due to the surface photovoltage (SPV).

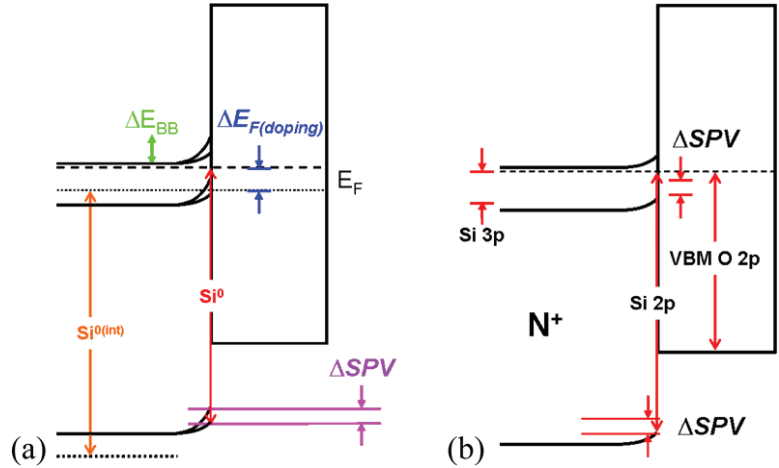

**Figure 6.** (a) Schema showing how doping, band bending and surface photovoltage interact to modify the electronic energy levels in $N^+$ zones of $N^+/P^-$ with respect to the values obtained for intrinsic silicon. The dotted lines represent intrinsic silicon; the dashed line is the position of the Fermi level in the heavily doped region. (b) Scale drawing of results for the $N^+$ region in $N^+/P^-$, with the calculated SPV effect.

**Table 2.** Calculated surface photovoltage effect following Hecht [17].

| Doping level (cm$^{-3}$) | $\Delta E_{SPV}$ (eV) |
|---|---|
| $10^{17}$ | $-0.4$ |
| $10^{20}$ | $-0.2$ |

**Table 3.** The estimated band bending at the oxide/substrate interface from the Si 2p analysis, corrected for the SPV effect.

| Pattern type | Band bending (eV) |
|---|---|
| $N^+$ | $-0.8$ |
| N | $-0.6$ |
| $P^+$ | $+0.5$ |
| $P^-$ | $+0.2$ |

Given the available photon energy range of the beamline and thus the inelastic mean free path, it was not possible to access the flat band region using synchrotron radiation. However, the binding energy for nearly intrinsic silicon is well-known and taken to be 99.35 eV [15]. $\Delta E_F^{doping}$ can be calculated using $N_D = n_i \exp((E_F - E_i)/kT)$ [16].

With synchrotron radiation, the high intensity induces a significant surface photovoltage caused by the creation of hole–electron pairs by the incoming photons. The local field caused by the upwards band bending (in the case of n type doping) at the substrate/oxide interface sweeps electrons into the bulk of the substrate and attracts holes in the valence band towards the interface. Thus the surface photovoltage acts to counter the band bending. In the case of p type doping the field is reversed, but the SPV acts again to 'flatten' the bands. The SPV can be calculated as [17]:

$$J = J_{\max}(1 - \exp(-\alpha W)) \approx J_0(T)\exp(-E_{CBO}/E_0).$$

The generated photocurrent $J$ is a function of the incoming flux, the photon absorption coefficient, the thermionic emission properties of the sample and the electron effective mass. We have calculated the $\Delta E_{SPV}$ for the CIPO flux of $10^{15}$ ph s$^{-1}$ cm$^{-2}$ as shown in table 2.

Thus the SPV, with the photon flux used on the CIPO beamline and the doping levels of the patterns and substrate, is far from being negligible, and can be of the same order of magnitude as the Fermi level shifts expected from the dopant concentration. The interaction of doping, band bending and surface photovoltage is shown in figure 6(a).

The results for the $N^+$ region are shown in figure 6(b) showing both core level and valence band binding energies. For $N^+/P^-$ the measured binding energy is 99.3 eV in the heavily n doped region, and 99.60 eV in the p doped substrate. As can be judged from equation (1), these values are the result of several contributions, and failure to take into account the different terms on the right-hand-side of equation (1) would inevitably result in a wrong interpretation of the band alignments. The effect of heavy n doping shifts the position of the Fermi level in the gap. With the low escape depth of the 2p electrons stimulated by the 127 eV incident photons, this is countered by the upwards band bending, which is itself partially flattened by the x-ray induced surface photovoltage. From the SPV values we can therefore deduce the real band bending at the oxide/substrate interface as a function of substrate doping. The results are summarized in table 3 for $N^+/P^-$ and $P^+/N$.

The spectroscopic imaging of $N^+/P^-$ using a monochromatic Al K$\alpha$ source gives an inelastic mean free path of $\sim$3 nm, making topmost surface effects negligible, whilst probing deeper into the substrate. The threshold image is shown in figure 7.

Although not carried out *in situ*, the results are valuable in checking the SPV calculation. The Si$^0$ binding energies determined using Al K$\alpha$ radiation are 99.1 eV ($N^+$) and 99.8 eV ($P^-$). In the case of the $N^+$ region this is just the value of the SPV, which is indeed expected to be much smaller, or even negligible with a laboratory source, compared to synchrotron radiation. In the limit of low flux, the SPV term in equation (1) is set to zero, thus the band bending as measured at the interface is simply equivalent in magnitude to the shift of the Fermi level due to doping. The values given in table 3 are indeed within 20% of the shifts expected from $\Delta E_F^{\text{doping}}$.

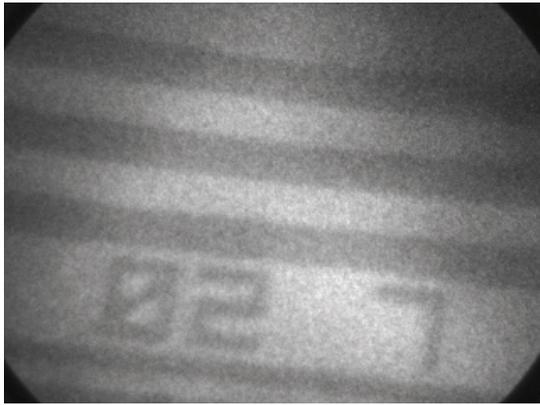

**Figure 7.** Threshold image obtained in laboratory conditions using an Al K$\alpha$ source and $E - E_F = 6.1$ eV. The FoV is 127 $\mu$m, and the acquisition time 8 s. The N$^+$ patterns are dark in the image.

Confirmation of the band bending at the substrate/oxide interface comes from the separation $\Delta\text{Si}^{4+}$ between the Si$^0$ and the Si$^{4+}$ components in the n and p type doped zones, clearly showing the effect of the opposite band bending expected for the two dopants, modulated of course by the SPV.

Thus, it is possible to deduce the spatial distribution of real band bending at the interface, free of effects induced by the photoemission analysis.

The O 2p bands of the SiO$_2$ native oxide are not bent; however, their alignment does follow the substrate valence bands. Thus the measured valence band onset for the O 2p levels reflects the combined effect of doping, shifting the Fermi level position in the gap and the substrate band bending. Thus the difference in valence band line-up is smaller than that expected from a flat band regime. The VBM of the Si 3p orbitals is not computed due to the low counting statistics extracted from the image series.

In the case of P$^+$/N, the statistics are better and the Si$^0$, Si$^{4+}$ binding energies can be quantified, as well as the valence band onsets of the O 2p and even the Si 3p orbitals. We recall that in the extracted Si 2p spectra in figures 3 and 4 the substrate signal was considerably sharper and more intense for P$^+$/N than for N$^+$/P$^-$. Most notable is the proximity of the measured binding energies in the n and p type regions, underlying the need for using quantitatively equation (1) in the interpretation of the experimental spectra.

The Si$^0$ binding energy measured from the 400 $\times$ 400 nm$^2$ AOIs can therefore be used to calculate the real band bending at the substrate/oxide interface as a function of doping type and level.

## 4. Conclusion

We have used fully energy-filtered x-ray photoelectron emission microscopy and synchrotron radiation in order to study variations in electronic band alignments at the native oxide/substrate interface as a function of p and n type doping in micron sized patterns on silicon wafers. High quality Si 2p spectra with good energy resolution (0.15–0.2 eV) are extracted from 400 $\times$ 400 nm$^2$ areas of interest and correlated with the valence band spectra from the same regions. Calculation of the surface photovoltage induced by the synchrotron radiation allows a direct estimate of the real band bending at the interface as a function of substrate doping. The results open perspectives for the study of multiple doping levels and doping gradients in silicon and other semiconductors and point to the use of energy-filtered XPEEM for non-destructive, but still offline, characterization of microelectronic devices.

## Acknowledgments

We thank the CIPO beamline scientists Nicola Zema and Stefano Turchini for their valuable help. This work was supported by the French ANR project 05-NANO-065 XPEEM.